\documentstyle[aps,preprint]{revtex}
\begin{document}
\draft
\title{\large
Low energy chaos in the Fermi-Pasta-Ulam problem}

\author{D. L. Shepelyansky$^{*}$}

\address {Laboratoire de Physique Quantique, UMR C5626 du CNRS, 
Universit\'e Paul Sabatier, F-31062 Toulouse Cedex 4, France}

\date{11 November 1996}

\maketitle

\begin{abstract}
A possibility that in the FPU problem the critical energy for chaos 
goes to zero with the increase of the number of particles in the chain
is discussed. The distribution for long linear waves in this regime is found
and an estimate for new border of transition to energy equipartition is given.
\end{abstract}

Starting from 1955 the Fermi-Pasta-Ulam (FPU) problem \cite{FPU} initiated 
numerous researches 
and became one of the corn stone in the modern statistical mechanics
\cite{LILI,JF}.
The absence of energy equipartition in the system of coupled nonlinear
oscillators observed numerically in \cite{FPU}
pushed forward the investigations of chaos as well as the
analysis of completely integrable systems (see  
\cite{JF} and refs. therein). 

The first explanation of the striking  result \cite{FPU}
was proposed by Chirikov and Izrailev \cite{CHIZ} on the
basis of the Chirikov criteria of overlapping resonances
\cite{1959}. According to \cite{CHIZ} it is necessary to exceed
some critical energy value to obtain
overlapping of the resonances, chaos  and
energy equipartition over linear modes. According to \cite{CHIZ}
in the case of low mode excitation 
(nonlinear sound waves) the critical
energy increases with the number of oscillators in the chain
(or energy per oscillator is constant). Below this energy 
it was argued that the resonances are not overlapped
and the motion is close to integrable one. Since some of the initial
conditions in \cite{FPU} were below this border 
the energy equipartition was absent \cite{CHIZ}. The 
results of \cite{CHIZ} were confirmed
in the series of analytical and numerical researches \cite{RUF}
where the authors also analyzed the dependence of the Lyapunov
exponents on the energy. However, these researches showed that the 
relaxation to an equilibrium distribution could be very
long at small energies that makes difficult to study
the transition from global chaos to intergable case.

In this paper the condition of resonance overlapping
for long (sound) waves in the limit of small energy is analyzed. 
For long waves the dispersion law is very close
to linear. Due to that for the system with finite but large number
of oscillators $N$ there are some terms in the nonlinear part of the
Hamiltonian which are in the resonance even for very low energies.
Such resonances being not considered in \cite{CHIZ} give a sharp
decrease of the chaos border in energy 
which goes to zero with the increase of the number of 
particles in the lattice. In this sense the 
long wave chaos can exist for arbitrary small nonlinearity.
The physical reason of this unusual phenomenon is connected with the
linearity of unperturbed system. Due to that for  
the sound dispersion
law which is typical for long waves the KAM theorem cannot be
applied and chaos can appear for arbitrarily small nonlinear
interaction. Such kind of phenomenon have been already studied in
different dynamical systems with few degrees of freedom 
\cite{JFPR,YM,QRD}. In such a case the dynamics can be described by 
a renormalized Hamiltonian independent on the strength of nonlinear
interaction. In particular the measure of chaotic component 
remains unchanged with the decrease of nonlinearity.
In the FPU-problem a deviation of the dispersion law from the linear
one gives rise to a critical chaos border which is, however,
extremely low and decreases with the number of particles in the chain.

We start our analysis from the $\alpha$-FPU problem with cubic
nonlinearity in the Hamiltonian:
\begin{equation}
H={1\over 2}{\sum_{n=0}^{N} [{p_{n}}^2 + {(x_{n+1}-x_{n})^2}]}+
{\alpha\over 3}{\sum_{n=0}^{N} {(x_{n+1}-x_{n})^3}}
\end{equation}
\indent
where the first sum gives the Hamiltonian $H_{0}$ of the linear waves 
and the second sum represents the interaction $H_{int}$. The boundary
conditions are fixed as $x_{0}=0; x_{N+1}=0$.
The eigenmodes $(Q_k,P_k)$ of 
$H_0$ are connected with the coordinates $x_n,p_k$ by the equations
 $x_{n}=\sqrt{2/(N+1)} \sum_{k} Q_{k} \sin(q_{k} n)$,
$p_n=\sqrt{2/(N+1)} \sum_{k} P_{k} \sin(q_{k} n) $
with $q_{k}=\pi k/(N+1)$, $1 \leq k \leq N$ \cite{LILI}. 
In this representation $H_0$ can be written as 
$H_0 = \sum_k {({P_{k}}^2+{\omega_k}^2 {Q_{k}}^2)/2}=
\sum_k {\omega_{k} I_{k}}$
with the eigenfrequencies $\omega_{k} = 2 \sin(q_k/2)$.
The action-angle variables $(I_k, \theta_k)$ are connected
with $(P_k,Q_k)$ in the standard way \cite{LILI}.

It is convenient to write the total Hamiltonian in the action-angle
variables $(I_k, \theta_k)$ of the linear problem. 
Taking into account that the nonlinear coupling 
is small we can keep in $H_{int}$ only the resonant terms
corresponding to the resonant 3-waves interaction. For long waves this 
condition corresponds to $k_3=k_2+k_1$. All other terms can be
eliminated by averaging over fast oscillations with frequencies
$\omega_k$. After this procedure we obtain the averaged Hamiltonian:
\begin{equation}
\bar{H}={\sum_{k} {\omega_{k} I_{k}}} +
{\alpha\over {2\sqrt{N+1}}}\sum_{k_1,k_2,k_3} 
(\omega_{k_1} \omega_{k_2} \omega_{k_3} 
I_{k_1} I_{k_2} I_{k_3})^{1/2} 
\cos(\theta_{k_3}-\theta_{k_2}-\theta_{k_1}) \delta_{k_3,k_1+k_2}
\end{equation}
\indent
which can be written as $\bar{H}=H_{0}+\bar{H_{int}}$. Here the bar marks 
the averaging over fast oscillations with 
$\omega_k \geq \omega_1=\pi/(N+1)$. The term fast means that
$\omega_1 \gg \delta\omega$ where $\delta\omega$ is the
typical nonlinear frequency  $\delta\omega  
\sim {\partial \bar{H_{int}}}/{\partial \theta} 
\sim \alpha {(E_{0}/N)}^{1/2}\omega_{k}$. For few low modes excited
around a given $k$-value we obtain $\delta\omega \sim \alpha (E_0/N)^{1/2}k/N$
where $E_0$ is initial energy. Following the way of \cite{CHIZ},
where the $\beta$-FPU model with quartic nonlinearity had been 
studied, we can find the chaos border from the condition
of the overlapping resonances $\delta \omega \sim \Delta \omega$
where $\Delta \omega \approx \omega_1 \approx \pi/N$ 
is the distance between the
main resonances in (1). According to this condition the
global chaos appears for 
$\tilde \alpha = \alpha {E_0}^{1/2} > \tilde \alpha_{CHI} \sim \sqrt{N}/k$. 

The Hamiltonian $\bar{H}$ has additional integral of motion
$E_{S}=\pi \sum_{k} k I_{k}/(N+1) \approx E_0$. 
For long sound waves ($k<<N$) we
can use approximate expression for the dispersion law
$\omega_{k}=q_{k}-{q_{k}^3}/24$ in $H_{0}$ while in the
term with $\bar{H_{int}}$ it is sufficient to use $\omega_{k}=q_{k}$.
By using the new resonant phases $\phi_{k}=\theta_{k}-q_{k}t$
we can transform (2) to the new resonant Hamiltonian:
\begin{equation}
{H_{R}}=-\sigma{\sum_{k=1}^{M} {k^{3} I_{k}}} +
2\mu\sum_{k_1=1}^{M} \sum_{k_2=1}^{M-k_1} 
(k_{1} k_{2} k_{k_{2}+k_{1}} I_{k_1} I_{k_2} I_{k_{2}+k_{1}})^{1/2} 
\cos(\phi_{k_{2}+k_{1}}-\phi_{k_2}-\phi_{k_1})
\end{equation}
\indent
where $\sigma=\pi^{3}/(24(N+1)^3)$,
$\mu=\pi^{3/2}\alpha/(4(N+1)^{2})$ and $M$ is the maximal number of
harmonics. It is convenient to introduce the new dimensionless time
$\tau = \mu t {\sqrt {E_{S}(N+1)/\pi}} $ in which the dynamics
is described by the renormalized resonant Hamiltonian 
\begin{equation}
\begin{array}{c}
{H_{RN}}=-\nu{\sum_{k=1}^{M} {k^{3} J_{k}}}\\
+2\sum_{k_1=1}^{M} \sum_{k_2=1}^{M-k_1} 
(k_{1} k_{2} ({k_{2}+k_{1}}) J_{k_1} J_{k_2} J_{k_{2}+k_{1}})^{1/2} 
\cos(\phi_{k_{2}+k_{1}}-\phi_{k_2}-\phi_{k_1})
\end{array}
\end{equation}
\indent
with one dimensionless parameter
\begin{equation}
\nu={{\sqrt{\pi}\sigma}\over{\mu \sqrt{(N+1)E_{S}}}}
={{\pi^2 } \over {6 \alpha \sqrt{E_{S}} (N+1)^{3/2}}}. 
\end{equation}
\indent
The new actions $J_k$ are connected with the old ones 
by the relation $I_k= E_{S} J_{k}(N+1)/\pi$. They are
now normalized by the condition $\sum_{k=1}^{M} k J_{k} =1$.

Let us analyze now the dynamics of the system (4). 
If initially only few modes are excited around a $k$-value
then the distance between the resonances is $\Delta \omega \approx \nu k^3$
while the width of the resonance is 
$\delta \omega \sim k^{3/2} {J_k}^{1/2} \sim k$.
From these estimates it is clear that the resonances are overlapped
\cite{1959} for $\nu < \nu_{cr} \sim 1/k^{2}$ and then chaos
arises. In the original variables this
means that the chaos border is given by
\begin{equation}
\alpha \sqrt{E_{S}} > \tilde \alpha_{s} \approx {k^{2} \over N^{3/2}}
\;\;\;\; or \;\;\;\;\; \nu < 1/k^{2}
\end{equation}
\indent
This border, which takes into account the degenerate
sound resonances $k_3=k_2+k_1$, 
decreases with the growth of $N$ and is $N^2$ times below the border
of global chaos $\tilde \alpha_{CHI}$. In the case of excitation of low modes
with $k \sim 1$ the critical energy above which the motion is chaotic
is $E_{c} \sim 1/(\alpha^2 N^3)$. Therefore, chaos arises at zero
temperature $T=E_{0}/N$.

For a better understanding of the properties of the system (4) a numerical 
investigations of its dynamical motion was carried out. The initial conditions
were usually fixed as three excited modes with $J_1=J_2=J_3=1/6$ 
and different phases $\phi$. The calculations of the maximal Lyapunov
exponent shows that above the border (6) the motion is characterized
by the positive exponent $\lambda_{RN}$ that indeed demonstrates 
the existence of chaos in this regime. Above the border the maximal Lyapunov
exponent is zero (except exponentially narrow chaotic layers).
A typical example of the dependence of  $\lambda_{RN}$ on 
renormalized time $\tau$ is presented in Fig.~\ref{f1}. The
energy distribution $E_k = k J_k$ over linear modes
is shown in Fig.~\ref{f2}. To suppress the fluctuations the values
of $E_k$ were averaged over time $\tau$ in the time interval
[1000-2000].
Below the chaos border (6) the number
of excited modes remains the same as for the initial distribution. 
On the contrary above this border the energy is distributed over
some finite width $\Delta k$ which is much larger than the initial width.
For the high values of $k \gg \Delta k$ the distribution decays 
in an exponential way. In the whole interval of $k$ the energy distribution
$E_{k}$ can be fitted by the effective distribution:
\begin{equation}
f_{k} = {A \over {l (\exp(k/l-\gamma)+1)}}
\end{equation}
\indent
where the length $l$ determines the effective number of excited modes, $\gamma$
is some constant which mainly effects the shape of the distribution for
small $k$ and $A$ is determined by $\gamma$ via the normalization
condition $\sum E_k \approx {\int f_k d k } =1 $. For the case of 
Fig.~\ref{f2}
the optimal value is $\gamma = 2.65$. 
It is interesting to note that the fitting (7) quite well
describes the distribution $E_k$ in the large
interval  $0.03<\nu<0.0005$ with the same $\gamma$
and different $l$. This fact is demonstrated in 
Fig.~\ref{f3} where 6 distributions are superimposed in the
rescaled variables $l E_k$ and $k/l$. 
The fitting (7) allows to determine the dependence
of the length $l$ on $\nu$. This dependence is presented in
Fig.~\ref{f4} and is approximately given by the  equation
$l= 0.42/\sqrt{\nu}$. The same 
functional dependence on $\nu$ takes place for
the quantity $1/E_{1}$ which characterizes the width of the distribution
$\Delta k$ for small $k$. The existence of the same scaling
on $\nu$ for $l$ and $1/E_1$ confirms once more that the
distribution $E_k$ has only one scaling parameter $l$.

The obtained scaling of $l$ from $\nu$ can be understood on the following
grounds. The nonlinear resonance width in (4)
is $\delta \omega \sim \partial H_{RN}/\partial J_{k} \sim
k^{3/2} \sqrt{J_{k}} k^{1/2}$ with $k \sim l$. The last term $k^{1/2}$ gives 
the result of summation over $k$ terms 
with random phases contributing in
$\delta \omega$. A typical distance between the
resonances is $\Delta \omega \sim \nu k^3$. The number of excited modes
is determined by the chaos border given by the resonance overlapping:
$\delta \omega > \Delta \omega$. According to this estimate
the number of excited modes is $\Delta k \sim l \sim 1/\sqrt{\nu}$
that is in agreement with the numerical dependence from Fig.~\ref{f4}
and the previous estimate (6). 
Using the expression for $\nu$ we can find  the 
effective number of excited linear modes expressed via the
original variables:
\begin{equation}
\Delta k \sim l \sim (\alpha^2 E_0 N^3)^{1/4}
\end{equation}
\indent
From this expression it follows that for fixed $\alpha$ and $E_0$
the number of excited modes is quite large but still
$\Delta k / N \ll 1$.  

In the same way we can obtain estimate for the maximal Lyapunov
exponent $\lambda_{RN}$ in the renormalized Hamiltonian (4). Indeed,
$\lambda_{RN} \sim \delta \omega \sim k^{3/2} \sqrt{k J_{k}} $ 
with $k \sim l$ and  $\lambda_{RN} \sim k \sim 1/\sqrt{\nu}$.
Using the relation between the time $t$ for the original system (1)
and the time $\tau$ in the renormalized Hamiltonian (4) we obtain the 
estimate for the maximal Lyapunov exponent $\Lambda$ in the system (1):
\begin{equation}
\Lambda = {{\pi \alpha \sqrt{E_S} \lambda_{RN}} \over {4 (N+1)^{3/2}}}
\sim {{\alpha^{3/2} {E_{0}}^{3/4}}\over {N^{3/4}}}
\end{equation}
\indent
The numerical data for the dependence of $\lambda_{RN}$ on $\nu$
are presented on the Fig.~\ref{f4}. 
Unfortunately, in the given interval of $\nu$
the variation of $\lambda_{RN}$ is not quite monotonic and
further numerical investigations are required for verification
of the theoretical dependence $\lambda_{RN} \sim 1/\sqrt{\nu}$
(see the discussion below).
Let us mention that the sign of $\nu$ in (4) is not important and 
the results are qualitatively the same
for $\nu<0$ when the absolute value of $\nu$ should be used in the estimates.

The comparison of $\Lambda$ with the distance between main resonances
$\Delta \omega$ shows that for sufficiently large $N$ the nonlinear resonance
width $\delta \omega \sim \Lambda $ becomes larger than 
$\Delta \omega \sim 1/N$. The condition 
$\Lambda >  \Delta \omega $ shows that the main resonances in (1) will be
overlapped for 
\begin{equation}
\alpha \sqrt{E_{o}} > \tilde \alpha_{eq} \approx 1/N^{1/6}
\end{equation}

Above this border the nonresonant terms neglacted in the derivation of (2)
give the overlapping of the main resonances and
for $\alpha E_0 > 1/N^{1/3}$  
approximate equipartition over all linear modes 
modes can be expected. So, in the limit of large $N$ the equipartition
can appear at zero energy and zero temperature. The time required to reach
the equipartition is inversely proportional to $\Lambda$.

It is interesting to note that some conditions of \cite{FPU} considered
usually as integrable (Fig.1 in \cite{JF}) have $\nu \approx 0.13$.
Direct computation in (4) for this $\nu$ value with 
corresponding initial conditions gives, however, $\lambda_{RN} =0$.
This puts the question about a more exact determination of the border
of chaos $|\nu_{cr}|$.

Let us now briefly discuss the properties of chaos in the
$\beta$-FPU model with quartic interaction 
$H_{int}=\beta \sum_n (x_{n+1} - x_{n})^4/4$. 
As in the $\alpha$-case we should keep only the
resonant terms for 4 waves with $k_1+k_2=k_3+k_4$. The 
resonance nonlinear width can be estimated in the same way as
in \cite{CHIZ,LILI}
$\delta \omega \sim \beta E_0 \omega_k/N$. The overlapping
of the main resonances happens for $\beta E_0 >N/k$ \cite{CHIZ,LILI}.
However, for the resonant Hamiltonian only the deviation of
$\omega_k$ from the sound law $\pi k /N$ is important so that
the distance between the resonances can be estimated as
$\Delta \omega \sim k^3/N^3$ \cite{REM}. This gives the
border of slow chaos $\beta E_0 > k^2/N$ which is much below
the standard border \cite{CHIZ,LILI}. Above this border the
number of excited low linear modes is 
$k \approx \Delta k  \sim \sqrt{\beta E_0 N}$ and the 
maximal Lyapunov exponent is 
$\Lambda \sim (\beta E_0/N)^{3/2} \sim \delta \omega$.
The overlapping of the main resonances takes place for $\Lambda > 1/N$
or $\beta E_0 > N^{1/3}$. Above this border all linear modes are
excited leading to energy equipartition. In a difference from
the $\alpha$-model this border grows with $N$ but the
critical temperature $T=E_0/N$ still goes to zero.

The above theoretical estimates were based on the 
comparison of the splitting between linear modes and
nonlinear spread width. As in the case of the Chirikov
criteria such approach cannot exclude a possibility
that the system under investigation is completely integrable
or is very close to some of them. This point is very
crucial for the $\alpha$-FPU problem since at low energy it is
very close to the Toda lattice (see \cite{LILI}).
Due to that generally we should expect that contrary to the above
estimates and numericaly data the dynamics of 
$\alpha$-FPU problem will be integrable. To understand this apparent
contradiction with the numerical data additional simulations 
had been carried out. Namely, the 
total number of harmonics $M$ had been increased up to 
$M=120$ for the parameters of Fig.1a. While the simulations become
very heavy in such a case they give approximately $\ln \tau/\tau$
decay of $\lambda_{RN}$ up to $\lambda \approx =0.02$ at 
maximally reached $\tau = 400$. This indicates that in a real
system with very large $M$ the Lyapunov exponent will be zero.
At the same time such change of $M$ did not affected the
averaged energy distribution (see Fig.2). For a better check of this 
point a number of numerical simulations with the original Hamiltonian
(1) had been done with $N$ up to 151 and the initial 
conditions corresponding to the Fig.1a with fixed $\nu=0.01$. 
For $ N=61$ the renormalized Lyapunov exponent (see (9))
was stabilized around $\lambda_{RN} \approx 0.13$
(the time $t_{max}$ in the simulations was $t_{max} \approx 9 \times 10^5$);
for $N=101$ the exponent was also stabilized around $\lambda_{RN} \approx
0.065 (t_{max} \approx 9 \times 10^6)$. However, in both these cases
the averaged energy distribution $E_k$ had significant increase 
at high modes in a difference from Fig.2. For $N=151$
during all $t < t_{max} \approx 6 \times 10^6$ the value
of $\lambda_{RN}$ was decreasing as $\ln \tau/\tau$ reaching
$\lambda_{RN} \approx 0.04$ at $t_{max}$. At the same time the averaged
distribution $E_k$ was practically the same as on the Fig.2
(see Fig.5).
These additional data show that in the low energy limit
the dynamics of $\alpha$-model is not chaotic ($\lambda_{RN}=0$)
as it can be expected from the comparison with the Toda lattice.
Inspite of that the energy distribution (see (7)) is correctly given
by the above estimates derived from the renormalized Hamiltonian.
The reason due to which the renormalized dynamics is so sensitive to
the maximal value of $M$ is still not quite clear.
It is possible that
the important effects of coupling to high modes can be understood
from nonlinear wave equation in the continuous limit
(see \cite{TUR}). 
Very recently, the properties of the Lyapunov exponent 
in the system (1) with N up to 128 were studied in \cite{PETTINI}.

The situation for the $\beta$-model can be more interesting.
Indeed, apparently this model is not close to any integrable system
and the above renormalization approach and estimates should
give correct chaos border. The picture of low energy
chaos developped here is qualitatively close to that one in 
\cite{REM1}. However, additional investigations of this regime 
are still highly desirable. They should 
clarify some uncertainties in the estimate of $\Delta \omega$
(see \cite{REM}). Also the question of coupling to high modes
can play a very crucial role \cite{RUFFO}. 

I would like to thank J.Bellissard and P.L'Eplattenier for stimulating
discussions. The above results had obtained at the begining of 1994.
I greatly appreciate the discussions and critisism of B.V.Chirikov
being in Toulouse at that time.


\figure{Fig. 1.
Maximal Lyapunov exponent $\lambda_{RN}$ in (4) as a function of time
$\tau$:
a)full line: $\nu = 0.01$, $H_{RN} = 0.544$, $\lambda_{RN}>0$;
b)dashed line: $\nu = 1.$, $H_{RN} = -5.45$, $\lambda_{RN} \rightarrow 0$
(values of $\lambda_{RN}$ are multiplied by 5). For all Figs. 1-4 
only three modes were initially excited with $J_1=J_2=J_3=1/6$.
\label{f1}}

\figure{Fig. 2.
Averaged energy distribution $E_k=k J_k$  over linear modes $k$
for the cases of Fig.1:
a) full circles; b) open circles. Full line gives the 
fitting distribution (7) with $l=4.05$; $\gamma=2.65$, $A=0.3678$.
\label{f2}}

\figure{Fig. 3.
 Normalized energy distribution $l E_k$ as function of $k/l$
for the six different values of $\nu$ from Fig.4 (points). The full
line gives the fitting (7) with $\gamma$ and  $A$ from Fig.2.
\label{f3}}

\figure{Fig. 4.
Dependence on $\nu$ for: length $l$ obtained from distributions of Figs. 3
(points); average energy of first mode $E_1$ (squares);
$\lambda_{RN}$ (open circles). The straight line 
shows the theoretical dependence $l \sim 1/\sqrt{\nu}$.
\label{f4}}

\figure{Fig. 5.
Same as Fig.2 obtained from the Hamiltonian (1) (see text).
\label{f5}}

\begin{thebibliography}{99}
\bibitem{REM}  Here for $\Delta \omega$ we used the 
               estimate $k^3/N^3$ which is correct for the majority
               of $k$. But for some $k$ values
               (e.g. near $k_1=k_2=k_3=k_4$) this distance
               is $k^2/N^3$ that gives 
               a wider spreading with $\Delta k \sim \beta E_0 N$,
               $\Lambda \sim (\beta E_0)^2/N$
               and equipartition border $\beta E_0 \sim 1$. 
               However, the chaos originated by these resonances should be 
               much slower.
\end{thebibliography}
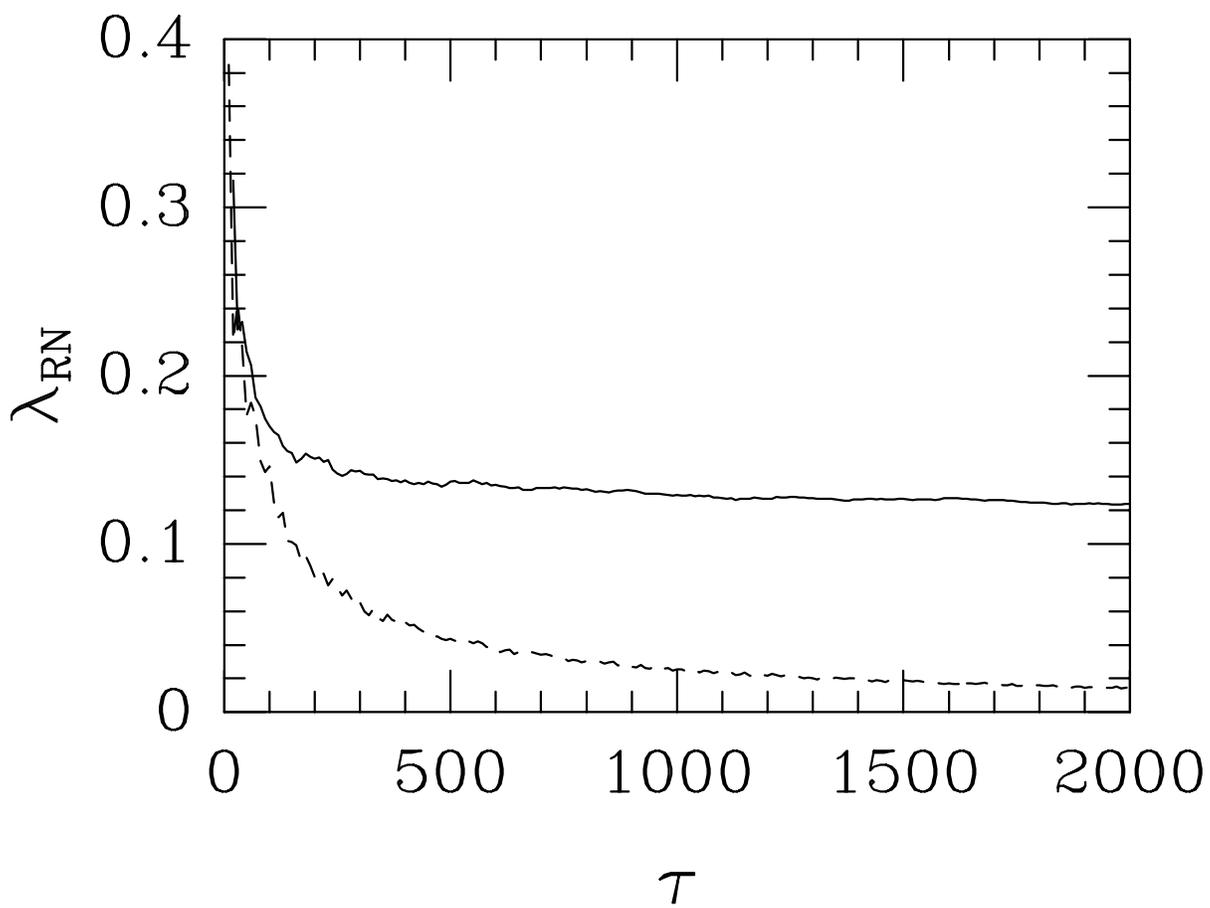
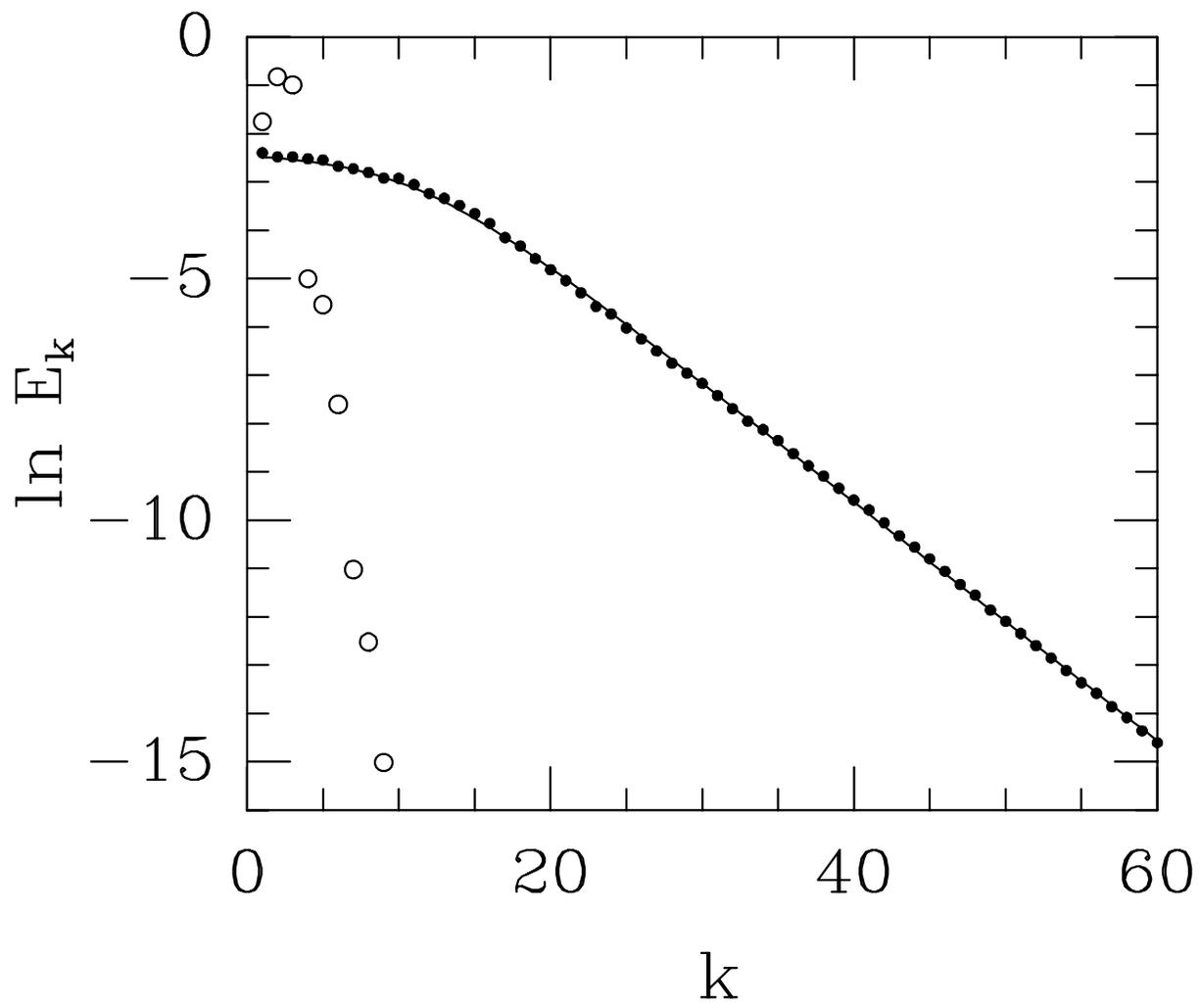
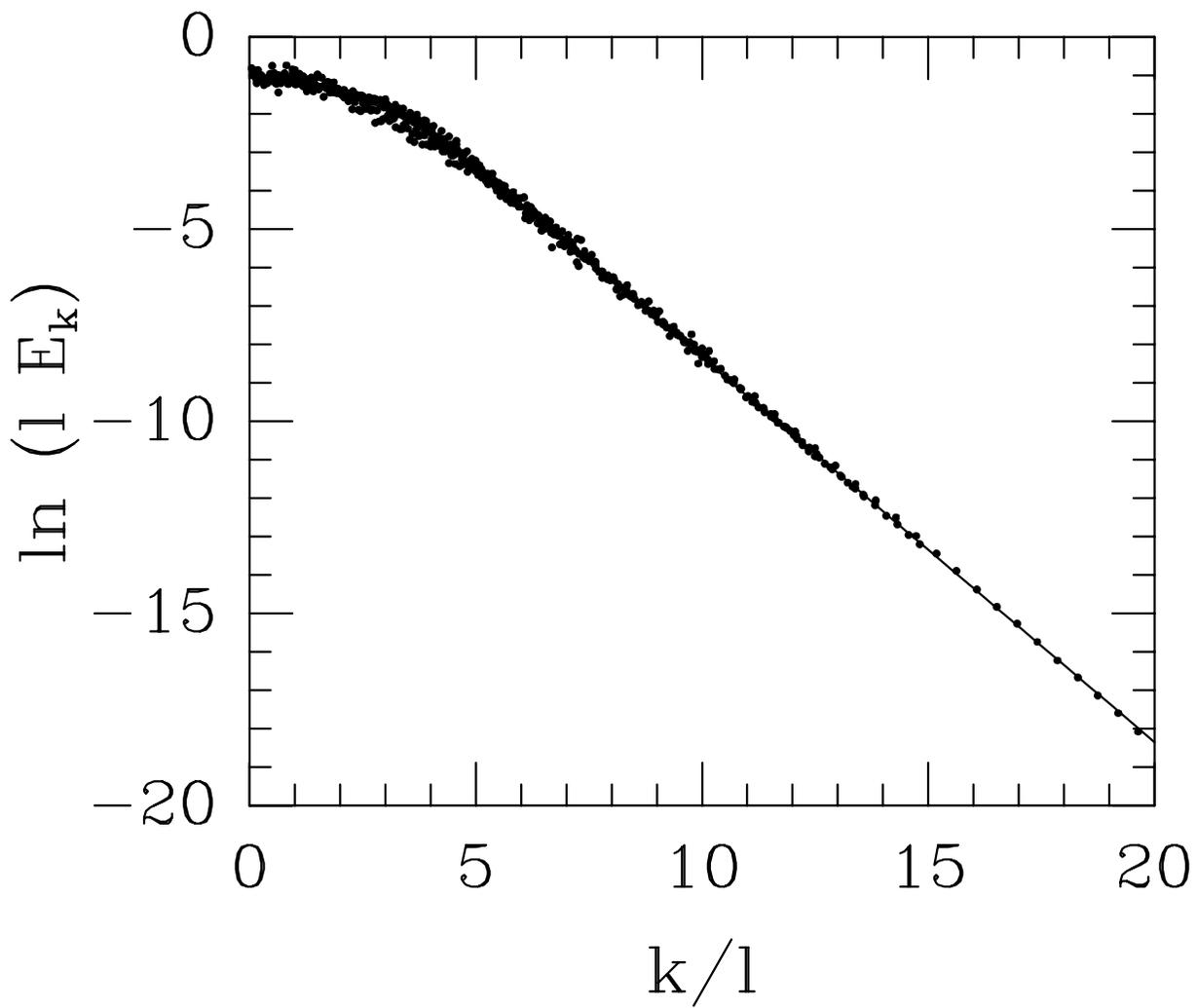
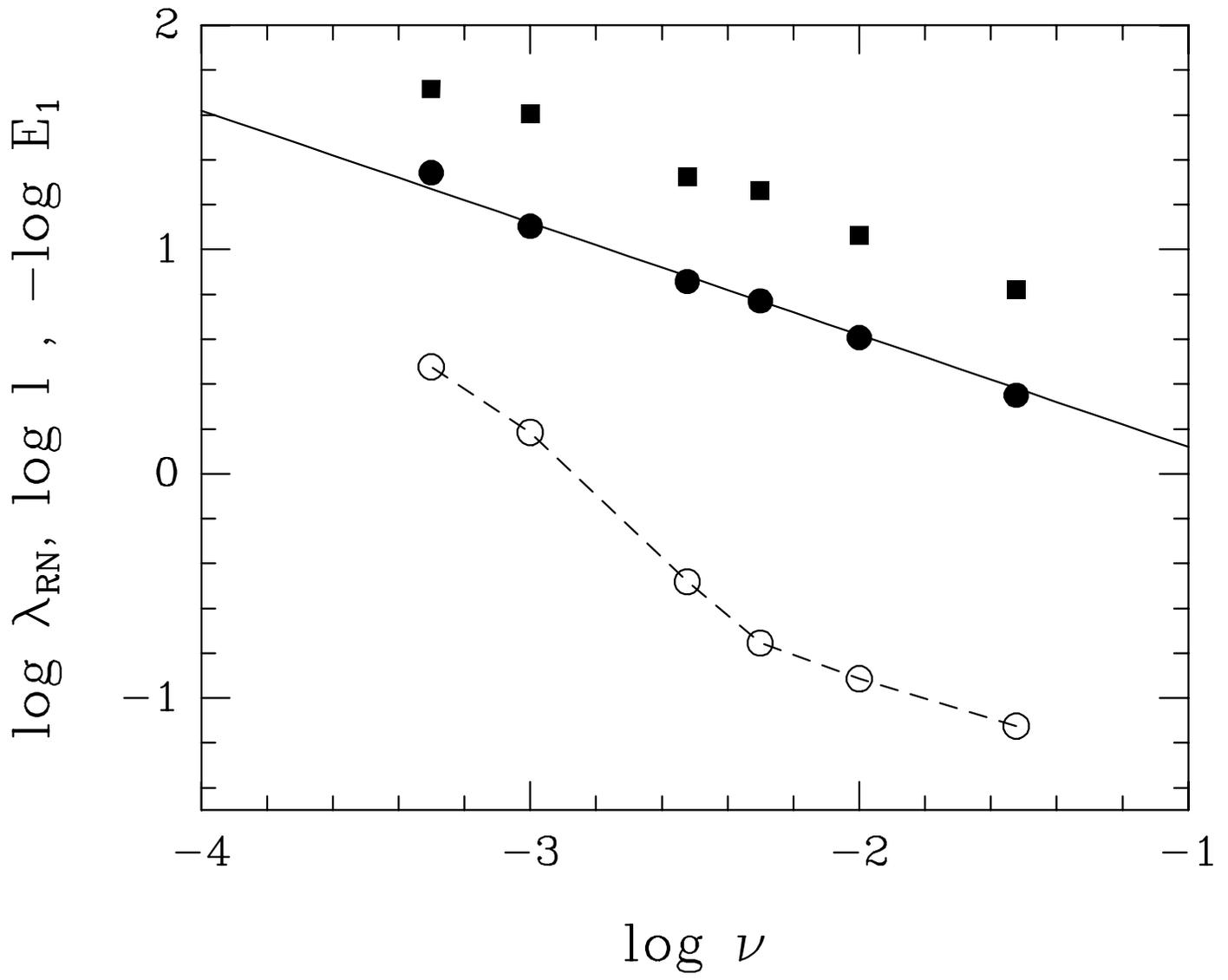
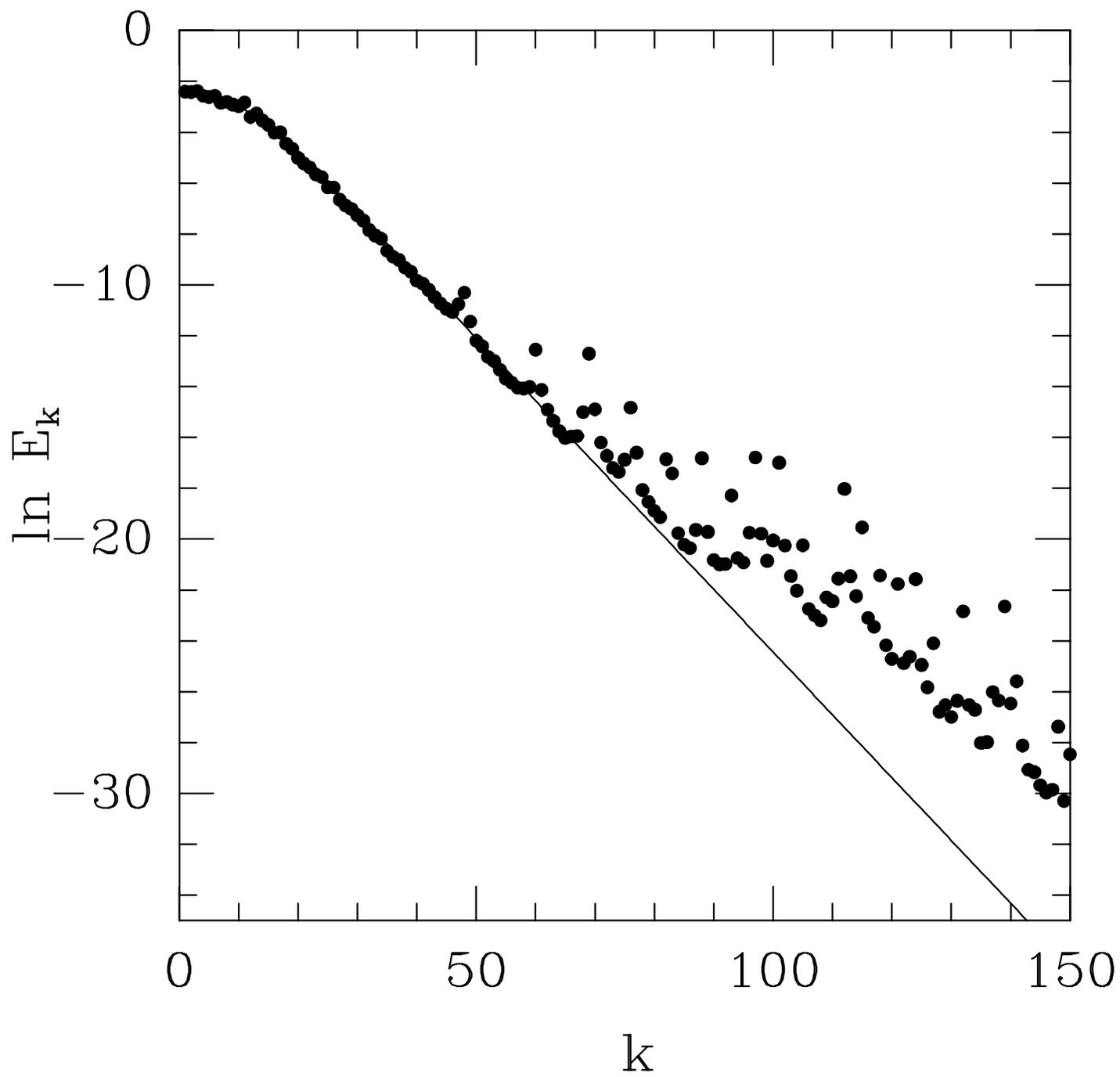
\end{document}